\documentclass[twocolumn,showpacs,preprintnumbers,amsmath,amssymb]{revtex4}

\usepackage{graphicx}

\begin{document}

\title{Accuracy of quantum-state estimation utilizing Akaike's information criterion}

\author{Koji Usami$^{1,2}$} \email{usami@frl.cl.nec.co.jp}
\author{Yoshihiro Nambu$^{2,3}$}
\author{Yoshiyuki Tsuda$^{4}$}
\author{Keiji Matsumoto$^{4}$}
\author{Kazuo Nakamura$^{1,2,3}$}  
\address{$^1$Depertment of Material Science and Engineering, Tokyo Institute of Technology,\\ 4259 Nagatsuta-cho, Midori-ku, Yokohama, Kanagawa, 226-0026,Japan}
\address{$^2$CREST, JST, 3-13-11 Shibuya, Shibuya-ku, Tokyo, 150-0002, Japan}
\address{$^3$NEC Fundamental Reseach Laboratories, 34 Miyukigaoka, Tsukuba, Ibaraki, 305-8501, Japan}
\address{$^4$ERATO, JST, 5-28-3 Hongo, Bunkyo-ku, Tokyo, 113-0033, Japan}

\date{\today}

\begin{abstract}
We report our theoretical and experimental investigations into errors in
quantum state estimation, putting a special emphasis on their asymptotic
behavior. Tomographic measurements and maximum likelihood estimation are
used for estimating several kinds of identically prepared quantum states
(bi-photon polarization states) produced via spontaneous parametric
down-conversion. Excess errors in the estimation procedures are eliminated
by introducing a new estimation strategy utilizing Akaike's information
criterion. We make a quantitative comparision between the errors of the
experimentally estimated states and their asymptotic lower bounds, which are
derived from the Cram\'{e}r-Rao inequality. Our results reveal influence of
entanglement on the errors in the estimation. An alternative measurement
strategy employing inseparable measurements is also discussed, and its
performance is numerically explored.
\end{abstract}

\pacs{03.67.-a, 42.50.-p, 89.70.+c}

\maketitle


\section{Introduction}

\label{sec:intro}

One of the central features of quantum mechanics is that it does not allow
to simultaneously obtain whole information about an individual quantum
system without errors \cite{MP1995}. The \textit{Holevo bound} on the
accessible information and the \textit{no-cloning theorem} are the prominent
manifestations of the restrictions on acquiring information from quantum
systems \cite{NCtext2000}, and these restrictions culminate in quantum
cryptography \cite{NCtext2000}.

However, there are no obstacles to estimate all aspects of quantum states in
a series of distinct measurements on identically prepared particles by 
\textit{quantum state tomography} \cite{NCtext2000,Leonhardt1997}. The
pioneering experimental demonstration of this method has been accomplished
by Smithey, \textit{et al.}~\cite{SBRF1993}. They determined a Wigner
function for vacuum and pulsed squeezed-vacuum state of a spatial-temporal
mode using \textit{homodyne tomography}. Schiller, \textit{et al.} \cite%
{SBPMM1996} applied this method to a estimation of a density matrix (in the
number state representation) for squeezed vacuum state of two spectral
components. In this experiment, the spectacular even-odd oscillations in the photon-number distribution was observed. Recently, Lvovsky \textit{et al.} %
\cite{LHABMS2001} and Bertet \textit{et al.} \cite{BAMOMBRH2002} have
respectively succeeded in reconstructing a Wigner function for single-photon Fock state of a travelling spatial-temporal mode and that of a
intra-cavity mode. Both estimated Wingner functions showed a dip reaching
classically-impossible negative values around the origin of the phase space.
For the polarization degree of freedom of electromagnetic field, White 
\textit{et al.} \cite{WJEK1999} used quantum state tomography, for the first
time, to characterize non-maximally entangled states produced from a
spontaneous-down-conversion photon source. Kwiat \textit{et al.} utilized
this method for the verification of decoherence-free characteristic of a
particular entangled state \cite{KBAW2000}, and for the demonstration of 
\textit{hidden} non-locality of entangled mixed states \cite{KBSG2001}.

In spite of these splendid experimental achievement with quantum state
tomography, statistical errors in estimating quantum states have been paid
minor attention so far. Statistical analyses of errors in quantum-state
estimation should not be undervalued. Since any outcomes of measurements are
represented as a random variable in quantum mechanics, statistical analyses
of their errors may reveal profound rule for acquiring information from
quantum system. Moreover, such analyses may also lead to the development of
quantum information technology, which requires us to faithfully prepare
several kinds of quantum states \cite{NCtext2000}, and to the improvement of the sensitivity for various kinds of precision measurements, which is limited by
quantum noises \cite{Phase,Polarization}.

In this article, we report our theoretical and experimental analyses of
errors in quantum state estimation putting a special emphasis on their
asymptotic behavior. In particular we focus on the estimation of the state
of two qubits (two 2-level quantum systems). The two-qubit system in
4-dimensional Hilbert space is the simplest one where the peculiar
characteristic of quantum mechanics, \textit{entanglement}, is activated.
Since entanglement plays the critical role in the mysterious phenomena in
the quantum world \cite{Bell,BWMEWZ1997,QCrypto2000}, it is interesting to
ask whether entanglement affects accuracy of the estimation. Various kinds
of two qubits (including entangled states) are practically realizable as polarization states of bi-photon produced via spatially-nondegenerate,
type-I spontaneous parametric down-conversion (SPDC) \cite%
{WJEK1999,KBAW2000,KBSG2001,KWWAE1999,JKMW2001,WJMK2001,NUTMN2002}. The
procedure to estimate the state of two qubits has been well established by
James, Kwiat, Munro and White \cite{JKMW2001}. Thus, in our experiments, we
followed the above methods for producing the ensembles of the bi-photon
polarization states, for measuring them, and for estimating their density
matrices.

The main purpose of this article is to quantitatively show the limit on
accuracy of quantum-state estimation. We demonstrate that the accuracy
depends on state to be estimated and also measurement strategy. In order to
do that, we introduce a new strategy of quantum-state estimation utilizing
Akaike's information criterion (AIC) \cite{Akaike1974} for eliminating
numerical problem in the estimation procedures especially in estimating
(nearly) pure quantum states. While number of parameters used for
characterizing density matrices of quantum states is fixed in the
conventional estimation strategies \cite{JKMW2001,BDPS1999,RHJ2001}, the
number is varied in the new strategy for eliminating redundant parameters.
Consequently, we can quantitatively compare experimentally-evaluated errors
in the estimation with their asymptotic lower bound derived from the Cram\'{e}%
r-Rao inequality without bothering about the delicate numerical problem
accompanying the redundant parameters. It is shown that the errors of the
experimental results nearly achieve their lower bounds for all quantum
states we examined. Moreover, owing to the reduction of the parameters, the AIC
based new estimation strategy makes the lower bounds slightly decreased.

Our results reveal that when measurements are performed locally (i.e.,
separately) on each qubit, existence of entanglement may degrade the
accuracy of estimation. Thus, while the measurements in our experiments are
local ones, we numerically examine the performance of an alternative
measurement strategy, which includes inseparable measurements on two qubits.

The remainder of the article is organized as follows. In Sec.~\ref%
{sec:experiment}, we show our experimental analyses of errors in estimating
density matrices as a function of the ensemble size, i.e., as varying data
acquisition time. In Sec.~\ref{sec:Bures}, we present a prescription for
calculating the asymptotic lower bounds on the errors in terms of fidelity
and show that in the asymptotic region, the errors should be decreasing as
inversely proportional to the ensemble size. Then we compare the lower
bounds with the experimental results. In Sec.~\ref{sec:AIC}, a new strategy
of quantum state estimation utilizing Akaike's information criterion is
introduced, and the accuracy of the state estimated by this new strategy is
presented. In Sec.~\ref{sec:collective}, the alternative measurement
strategy for two qubits, which employs inseparable measurements, is
numerically explored. Section~\ref{sec:conclusion} summarizes this article.
In the Appendix, we briefly review tomographic measurements and maximum
likelihood estimation for estimating two qubits, and derive the Cram\'{e}%
r-Rao lower bound on the errors in the estimation.


\section{experiment}

\label{sec:experiment}


\subsection{experimental setup}

\label{subch:setup}

For experimentally producing various quantum states of two qubits, we use
the method to create the various polarization states of bi-photon via
spatially-nondegenerate, type-I spontaneous parametric down-conversion
(SPDC). The method was invented by Kwiat \textit{et~al.} \cite{KWWAE1999}
and applied to the various experiments \cite%
{WJEK1999,JKMW2001,KBAW2000,KBSG2001,WJMK2001,NUTMN2002}.

\begin{figure}[tbp]
\includegraphics[width=\linewidth]{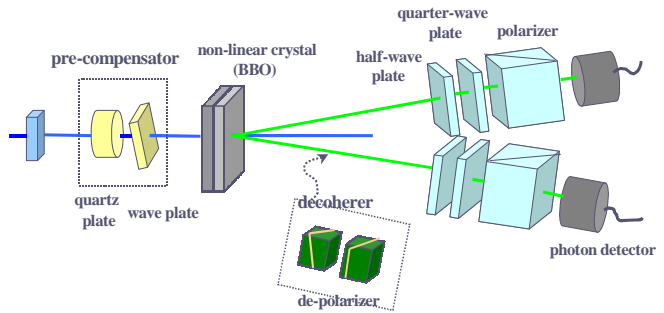}
\caption{Experimental setup for producing various polarization states of
bi-photon and measuring them.}
\label{fig:setup}
\end{figure}

A rough sketch of the experimental setup is shown in Fig.~\ref{fig:setup}.
Two thin (0.13mm) beta-barium borate ($\beta $-$BaB_{2}O_{4}$, BBO)
crystals, which are cut for satisfying the type-I phase matching, are
adjacent so that their optical axes lie in the planes perpendicular to each
other. Inside the crystals, the third harmonic beam (wave length:~266nm,
average power:~190mW) of the mode-locked Ti:Sapphire laser (pulse
duration:~80fs, repetition rate:~82MHz) -we will call it pump beam- is
slightly converted into the frequency-degenerate, but
spatially-nondegenerate (opening angle: $3^{\circ}$) bi-photon (wave
length:~532nm) via SPDC. This configuration of the setup makes
it possible to produce various polarization states of bi-photon (including
entangled states) by adjusting the pump beam polarization with a half-wave
plate (HWP) \cite{WJEK1999,KWWAE1999}, by modifying the relative time delay
between the horizontal and vertical components of the pump beam with a 
\textit{pre-compensator} (which consists of quartz plates and a variable
wave plate (WP)) \cite{NUTMN2002}, and by inserting \textit{decoherers} (two de-polarizers) into one of the paths of the down-converted photons \cite%
{KBAW2000,KBSG2001,WJMK2001}, as shown in Fig.~\ref{fig:setup}. We produced
three particular quantum states, the \textit{very noisy mixed state (VNMS)},
the \textit{almost pure and separable state (APSS)}, and the \textit{highly
entangled state (HES)} for inspecting influence of the various
characteristics (e.g., entropy and entanglement) of the states on the
accuracy of the estimation.

The produced polarization states of bi-photon were estimated by tomographic
measurements \cite{JKMW2001,WJEK1999} and maximum likelihood estimation
(MLE) \cite{JKMW2001,BDPS1999,RHJ2001}. These procedures are reviewed in
Appendices~\ref{app:tomography} and \ref{app:MLE}. In tomographic
measurements, the coincidental detection events (within 6ns) on both
single-photon detectors (HAMAMATSU H7421-40) were counted by using the time
interval analyzer (YOKOGAWA TA-520) during the data acquisition time $t$ at
each polarizer's setting (i.e., projector) $|m_{\nu }\rangle \langle m_{\nu
}|$ (which was determined and varied by the half-wave plate (HWP), the
quarter-wave plate (QWP), and the polarizer (Pol) on each path of the produced photons).
For investigating ensemble size-dependence of the accuracy, we varied the
data acquisition time of each measurement $t$ as 0.2s, 0.5s, 1.0s, 2.0s, and
5.0s. The typical single counting rate was about 30000c/s with the dark
counting rate of about 300c/s. The typical coincidence counting rate was
roughly 500c/s with the accidental coincidence counts being below 1\% of the
genuine coincidence counts. For eliminating the ambient photons, we used the
interference filters (FWHM: 8nm) (see Ref.~\cite{NUTMN2002}, for more
detailed information).


\subsection{experimental procedure}

\label{subch:procedure}

In order to assess the accuracy of the estimation, we repeated the
measurements and estimation procedures 9 times for each state and each
ensemble size. Here, as noted in Appendix~\ref{app:MLE}, the density matrix
of the two-qubit(2 two-level quantum state) can be written as 
\begin{equation}
\rho _{\Theta}=\frac{T_{\Theta}T_{\Theta}{}^{\dagger}}{\mathop{\mathbf{Tr}} 
\nolimits[T_{\Theta}T_{\Theta}{}^{\dagger}]},  \nonumber
\end{equation}
which satisfies the positivity condition and the trace condition for density
matrices \cite{JKMW2001,BDPS1999}; see Appendix \ref{app:MLE}. As a result
of the 9 identical trials, we had 9 slightly different density matrices $%
\{\rho_{\hat{\Theta_{i}}}\}_{i=1}^{9}$. The differences of these
states might stem not only from the statistical errors but also from the
experimental systematic ones. For reducing the systematic errors, we
restricted our data acquisition time, $t$, at each polarizer setting up to $%
t=5s$, so as to keep the experimental condition unchanged (especially, to
keep the pump power constant during whole data acquisition time $t \times 16$
measurements).

Then we evaluated the accuracy of the estimation in terms of the average 
\textit{fidelity} between the \textit{true} state $\rho_{\Theta_{0}}$ and
each estimated state $\{ \rho_{\hat{\Theta_{i}}}\}_{i=1}^{9}$, i.e., 
\begin{equation}
F(\rho_{\Theta_{0}},\rho_{\hat{\Theta}}) \approx \frac{1}{9}%
\sum_{i=1}^{9}F(\rho_{\Theta_{0}},\rho_{\hat{\Theta_{i}}}),
\label{eq:averageF}
\end{equation}
where the fidelity $F(\rho_1,\rho_2)$ is equal to $\mathop{\mathbf{Tr}}%
\nolimits[\sqrt{\sqrt{\rho_1}\rho_2\sqrt{\rho_1}}\,]$ \cite%
{NCtext2000,Jozsa1994,BCFJS1996}. As the \textit{true} state $%
\rho_{\Theta_{0}}$ for each of our concerned three states (the VNMS, the
APSS, and the HES), we employed a state which was estimated by the MLE using
the \textit{whole} data acquired for each state. This means that the
effective data acquisition time for determining the \textit{true} state
amounts to $t$=0.2s$\times$9-trial+0.5s$\times$9-trial+1.0s$\times$%
9-trial+2.0s$\times$9-trial+5.0s$\times$9-trial=78.3s. Later on we use these
three \textit{true} states as sources to produce artificial 16
coincidence-count data for the numerical simulations. These simulations are
performed without considering any systematic errors. Thus we can evaluate to
what extent the systematic errors affect the total errors.


\subsection{experimental results}

\label{subch:results}

\begin{figure*}[tbp]
\includegraphics[width=0.7\linewidth]{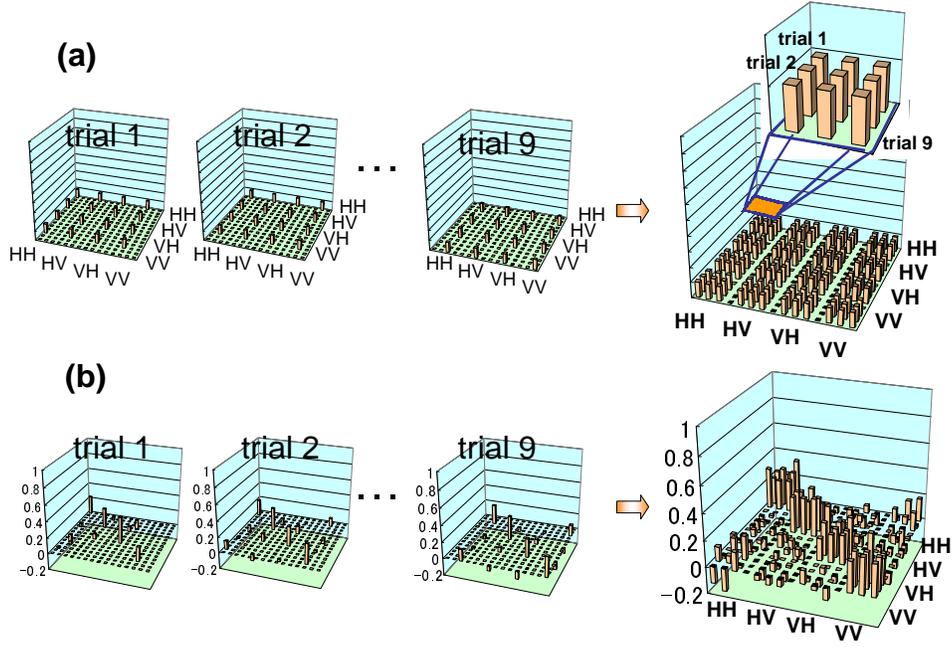}
\caption{To see fluctuation of the estimation, we use a matrix shown in the
most right-hand side of (a). This matrix has $12\times12$ elements and is
composed of 9 matrices ($4\times4$ matrix) shown in the left-hand side of
(a). Each of the 9 matrices is the real part of the density matrix estimated
by each trial. The relation between the elements of $12\times12$ matrices
and those of the $4\times4$ matrices is illustrated in (a). As an example,
the $12\times12$ matrix and the constituent $4\times4$ density matrices for
the VNMS are shown in (b). The bases of the density matrices are $|HH\rangle 
$, $|HV\rangle $, $|VH\rangle $ and $|VV\rangle $ (these notations are
defined in Appendix~\ref{app:tomography}). In this way, fluctuations of
estimated density matrices can be visualized.}
\label{fig:matrix}
\end{figure*}

\begin{figure*}[tbp]
\includegraphics[width=0.7\linewidth]{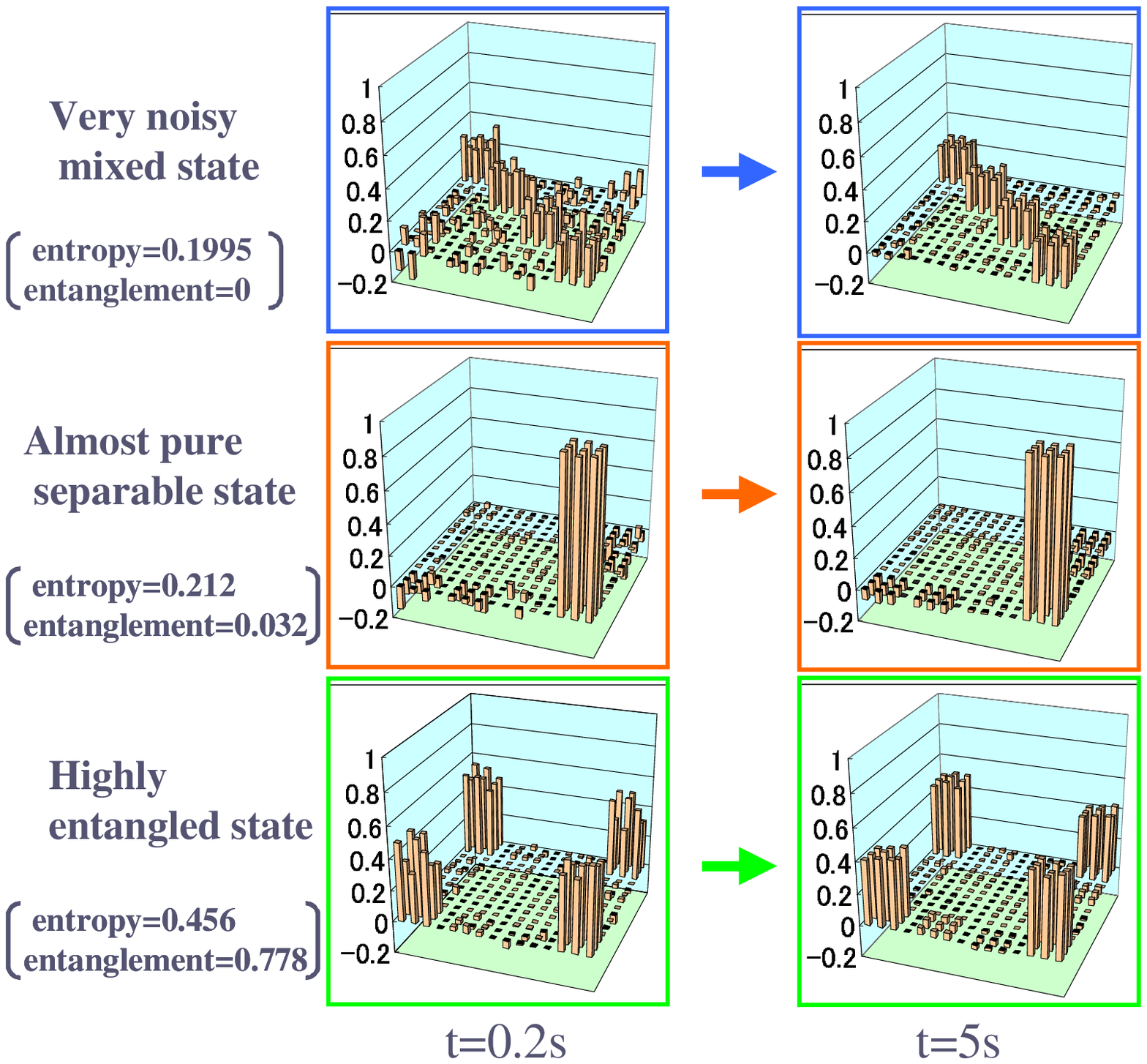}
\caption{The $12\times12$ matrix (which is composed of 9 density matrices ($%
4\times4$)) for the three states, i.e., the VNMS, the APSS, and the HES, for
two different data acquisition time, $t=0.2s$ (left) and $t=5.0s$ (right).
The detailed explanation for each element of the matrices is provided in
Fig.~\ref{fig:matrix}. Entropy (von Neumann entropy) and entanglement
(entanglement of formation) of each resulting \textit{true} state (which is
obtained by using the \textit{whole} data as mentioned in the text) is also
presented. Note that only the real parts of the density matrices are
exhibited, because the imaginary parts are quite small compared with the
real parts.}
\label{fig:devmatrix}
\end{figure*}

To visualize fluctuation of the estimation, we use a matrix, which is
explained in detail in Fig.~\ref{fig:matrix}. This matrix has $12\times12$ elements
and is composed of the 9 matrices ($4\times4$ matrix), each of which is the
real part of the density matrix estimated by each trial. Figure~\ref%
{fig:devmatrix} shows the matrices for three states (i.e., the VNMS, the
APSS, and the HES) for two different data acquisition time, $t=0.2s$ and $%
t=5.0s$. Here the bases of the density matrices are $|HH\rangle $, $%
|HV\rangle $, $|VH\rangle $ and $|VV\rangle $ (these notations are defined
in Appendix \ref{app:tomography}). Entropy (von Neumann entropy \cite%
{NCtext2000}) and entanglement (entanglement of formation \cite{NCtext2000})
of each resulting \textit{true} state, which is obtained by using the 
\textit{whole} data as mentioned before, is also shown in Fig.~\ref%
{fig:devmatrix}. Note that the each density matrices had little imaginary
parts in all cases, thus they are not presented. We can observe that the
fluctuation of each element is reduced as the data acquisition time, $t$,
becomes long (from 0.2s to 5.0s) in all three states.

\begin{figure}[tbp]
\includegraphics[width=0.95\linewidth]{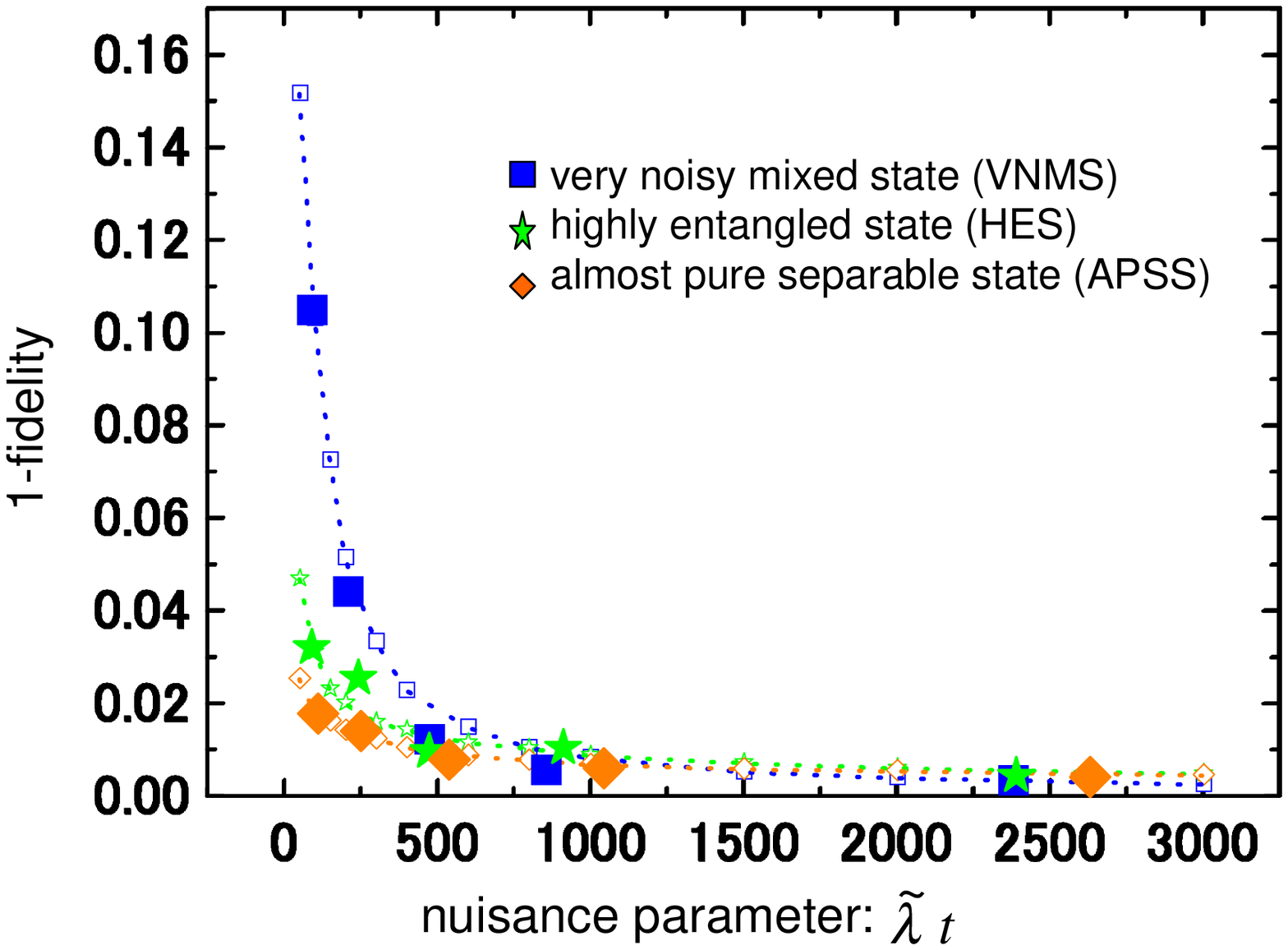}
\caption{The errors in estimating the three states (the VNMS, the APSS, and
the HES) in terms of fidelities between the \textit{true} states and the
estimated states are shown as a function of the nuisanece parameter, $\tilde{%
\protect\lambda}\,t$ ($\tilde{\protect\lambda} \approx 500$ is coincidence
counting rate and $t$ is the data acquisition time of each measurement). The
filled plots represent experimental results (the error bars are omitted) and
the blank plots represent numerical simulations.}
\label{fig:result}
\end{figure}

In Fig.~\ref{fig:result}, the fluctuations of the estimated density matrices
are quantitatively shown in terms of the average fidelities between the 
\textit{true} state $\rho_{\Theta_{0}}$ and the estimated states $\{\rho_{%
\hat{\Theta_{i}}}\}_{i=1}^{9}$ as a function of the ensemble size. The ensemble size corresponds to the nuisance parameter of the estimation, $\lambda =\tilde{\lambda}\,t$, where $\tilde{\lambda}\approx 500$ is coincidence counting
rate and $t$ is the data acquisition time of each measurement (see Appendix %
\ref{app:tomography} for the detailed explanation). Each filled plot
corresponds to the experimental result of the average fidelity, Eq.~(\ref%
{eq:averageF}). To supplement the experiments, we also carried out Monte
Carlo simulations by artificially producing 16 coincidence-count data, $%
\{N\} $, according to their \textit{true} states $\rho_{\Theta_{0}}$ and to
the probability mass function given by Eq.~(\ref{eq:pdf}). The estimation
procedures are the same as the experiments except for no systematic errors.
These simulations were repeated 200 times, therefore, the blank plots
correspond to 
\begin{equation}
F(\rho _{\Theta _{0}},\rho _{\hat{\Theta}})\approx \frac{1}{200}%
\sum_{i=1}^{200}F(\rho _{\Theta _{0}},\rho _{\hat{\Theta _{i}}}).
\label{eq:averageF_U}
\end{equation}
Note that the repetition of the simulations (200 times) may be large enough
to ensure their statistical confidence. The results of the numerical
simulations are in good agreement with those of the experiments. Thus the
systematic errors seem to be negligible for our experimental condition
(i.e., for the relatively short data-acquisition time). Nevertheless, we
remark some sources of our systematic errors; the fluctuation of the pump
power for SPDC during the 16 measurements (about 1.0\%), the uncertainty of
the wave-plate's angular setting in the tomographic measurements (about $%
0.05^{\circ}$), the finite extinction ratio of the polarizers (about $%
\frac{1}{400}$), and the accidental coincidence counts (about 1\% of the
genuine coincidence counts).

\section{accuracy of quantum state estimation}

\label{sec:Bures}

Suppose that there are two quantum states, $\rho_{1}$, and $\rho_{2}$; how
can the distance between these two quantum states be measured? One possible
answer is known as the \textit{Bures distance} \cite%
{Jozsa1994,BC1994,Holevo2001}: 
\begin{eqnarray}
d_{Bures}(\rho _{1},\rho _{2})^{\,2} &=&2(1-\mathop{\mathbf{Tr}}\nolimits[%
\sqrt{\sqrt{\rho _{1}}\rho _{2}\sqrt{\rho _{1}}}\,])  \nonumber \\
&=&2(1-F(\rho _{1},\rho _{2})),  \label{eq:Bures}
\end{eqnarray}
where, $F(\rho _{1},\rho _{2})$ is the fidelity. In Sec.~\ref{sec:experiment}, we have already evaluated the accuracy of the estimation in terms of the fidelities between the \textit{true} state, $\rho _{\Theta_{0}}$, and the estimated states, $\rho_{\hat{\Theta}}$. We will derive the highest accuracy, which is, in principle, attainable by our tomographic measurements in terms of the Bures distance. Then, it is compared with the experimental results.

Assuming that the estimated states $\rho_{\hat{\Theta}}$ are in the neighborhood of the \textit{true} state $\rho_{\Theta_{0}}$, the average Bures distance between them can be written as \cite{BC1994,Holevo2001,Uhlmann1993,FN1995,FujiwaraPhD,MatsumotoPhD,Hayashi2002} 
\begin{eqnarray}
&{}&d_{Bures}(\rho _{\Theta _{0}},\rho _{\hat{\Theta}})^{\,2}  \nonumber \\
&\approx &\!\!\frac{1}{4}\sum_{i=1}^{16}\sum_{j=1}^{16}J_{ij}^{SLD}(%
\Theta_{0})\,(\hat{\theta}^{i}(N)-\theta _{0}^{i})(\hat{\theta}%
^{j}(N)-\theta_{0}^{j}),  \label{eq:Riemann}
\end{eqnarray}
where $\{\theta _{0}^{i}\}_{i=1}^{16}\equiv \{\Theta _{0}\}$ are the \textit{%
true} parameters characterizing the \textit{true} state $\rho_{\Theta_{0}}$ %
\cite{note_Bures}, and $\{\hat{\theta}^{i}(N)\}_{i=1}^{16}$ are their
estimates inferred from the results of tomographic measurements $\{N\}\equiv
\{n_{\nu }\}_{\nu =1}^{16}$; see Appendix \ref{app:tomography} and \ref{app:MLE}. Here $[J_{ij}^{SLD}(\Theta )]\equiv \mathop{\mathbf{J}}%
\nolimits^{SLD}(\Theta)$ is a $16\times16$ matrix given by the following manner. First, we define a Hermitian operator $\mathop{\mathbf{L}}\nolimits_{i}^{S}(\Theta )$ called \textit{symmetric logarithmic derivative (SLD)} \cite{Holevo2001,Helstrom1976,Holevo1982,ANtext2000}, by 
\begin{equation}
\frac{\partial \rho _{\Theta }}{\partial \theta ^{i}}=\frac{1}{2}(%
\mathop{\mathbf{L}}\nolimits_{i}^{S}(\Theta )\,\rho _{\Theta }+\rho
_{\Theta}\,\mathop{\mathbf{L}}\nolimits_{i}^{S}(\Theta )).  \label{eq:SLD}
\end{equation}
The SLD $\mathop{\mathbf{L}}\nolimits_{i}^{S}(\Theta)$ can be obtained by
solving the equation above and considered as a quantum analogue of the 
\textit{score} (classically, the score is defined by $\frac{\partial}{\partial \theta^{i}}\ln[P(N|\Theta)]$ as noted in Appendix \ref{app:CR bound}). Then the matrix, $\mathop{\mathbf{J}}\nolimits^{SLD}(\Theta)$, called \textit{symmetric logarithmic derivative Fisher information matrix (SLD Fisher information matrix)}, is given by 
\begin{equation}
J_{ij}^{SLD}(\Theta)=\frac{1}{2}\mathop{\mathbf{Tr}}\nolimits[\rho
_{\Theta}(\mathop{\mathbf{L}}\nolimits_{i}^{S}(\Theta )\mathop{\mathbf{L}}%
\nolimits_{j}^{S}(\Theta )+\mathop{\mathbf{L}}\nolimits_{j}^{S}(\Theta )%
\mathop{\mathbf{L}}\nolimits_{i}^{S}(\Theta ))].  \label{eq:SLDFisher}
\end{equation}

From Eq.~(\ref{eq:Riemann}), we can see that the Bures distance is locally
equivalent to a distance on a Riemannian manifold equipped with a metric
structure defined by the SLD Fisher information matrix \cite%
{BC1994,Holevo2001,Uhlmann1993,FN1995,FujiwaraPhD,MatsumotoPhD,Hayashi2002}.
This recognition furnishes us with a geometrical picture of quantum state
estimation.

We note here that the SLD Fisher information matrix Eq.~(\ref{eq:SLDFisher})
was originally introduced for extending classical parameter estimation
theory to its quantum counterpart and formulating the \textit{quantum} Cram%
\'{e}r-Rao type lower bound on the errors in estimating quantum states \cite%
{Holevo2001,Helstrom1976,Holevo1982,ANtext2000,YL1973,BC1994,FN1995,FujiwaraPhD,MatsumotoPhD,Hayashi1998,FN1999,GM2000,Matsumoto2002}.

Our aim here is to evaluate the best accuracy in estimating identically
prepared quantum states $\rho_{\Theta_{0}}$ by tomographic measurements. In
other words, our aim is to find out the minimum Bures distance between the 
\textit{true} states and the estimated states, which can be attained by our
tomographic measurements. This can be accomplished by decreasing the value $(%
\hat{\theta}^{i}(N)-\theta_{0}^{i})(\hat{\theta}^{j}(N)-\theta _{0}^{j})$ of
Eq.~(\ref{eq:Riemann}) as much as possible. As is derived in Appendix \ref%
{app:CR bound}, the lower bound on the covariance $\mathop{\mathbf{E}}%
\nolimits_{\Theta_{0}}[(\hat{\theta}^{i}(N)-\theta_{0}^{i})(\hat{\theta}%
^{j}(N)-\theta _{0}^{j})]$ can be obtained by the \textit{Cram\'{e}r-Rao
inequality} 
\begin{eqnarray}
\mathop{\mathbf{E}}\nolimits_{\Theta_{0}}[(\hat{\theta}^{i}(N)-%
\theta_{0}^{i})(\hat{\theta}^{j}(N)-\theta _{0}^{j})] &\equiv&
V^{ij}(\Theta_{0})  \nonumber \\
&\ge& J_{ij}^{\,-1}(\Theta_{0}),  \label{eq:CRelement}
\end{eqnarray}
where $\mathop{\mathbf{E}}\nolimits_{\Theta_{0}}[f(N)]$ means averaging over
the \textit{true} probability mass function of $\{N\}$ given by Eq.~(\ref{eq:pdf}%
) with $\Theta=\Theta_{0}$; $\mathop{\mathbf{J}}\nolimits_{ij}(\Theta_{0})$
is the \textit{Fisher information matrix} (see Appendix~\ref{app:CR bound}).
Since the lower bound on the covariance is known to be asymptotically
achievable by using the MLE, the achievable lower bound on the Bures
distance can be given by 
\begin{eqnarray}
d_{Bures}(\rho _{\Theta _{0}},\rho _{\hat{\Theta}})^{\,2} &\approx &\frac{1}{%
4}\sum_{i=1}^{16}\sum_{j=1}^{16}J_{ij}^{SLD}(\Theta _{0})\,V^{ij}(\Theta_{0})
\nonumber \\
&\geq &\frac{1}{4}\mathop{\mathbf{Tr}}\nolimits[\mathop{\mathbf{J}}%
\nolimits^{SLD}(\Theta _{0})\,\mathop{\mathbf{J}}\nolimits^{-1}(\Theta_{0}].
\label{eq:limitTM}
\end{eqnarray}

From Eqs.~(\ref{eq:nuisance}), (\ref{eq:pdf}), and (\ref{eq:Fisher2}), the
Fisher information matrix $J_{ij}(\Theta_{0})$ can be rewritten as 
\begin{equation}
J_{ij}(\Theta_{0}) \approx \lambda\bar{J_{ij}}(\Theta_{0})=t\tilde{\lambda}%
\bar{J_{ij}}(\Theta_{0}),  \label{eq:Timprove}
\end{equation}
where 
\begin{equation}
\bar{J_{ij}}(\Theta_{0}) = -\mathop{\mathbf{E}}\nolimits_{\Theta_{0}} [\frac{%
\partial^{2}}{\partial \theta^{i} \partial \theta^{j}} \ln [\bar{P}%
(N|\Theta)]|_{\Theta=\Theta_{0}}],  \label{eq:scaledFisher}
\end{equation}
with 
\begin{equation}
\bar{P}(N|\Theta) = \prod_{\nu=1}^{16}e^{-\mathop{\mathbf{Tr}}\nolimits[%
|m_{\nu} \rangle\langle m_{\nu}|\ \rho_{\Theta}]} \frac{\mathop{\mathbf{Tr}} 
\nolimits[|m_{\nu} \rangle\langle m_{\nu}|\ \rho_{\Theta}]\,^{n_{\nu}}}{%
n_{\nu}!}.  \label{eq:scaledPDF}
\end{equation}
Therefore, 
\begin{equation}
\mathop{\mathbf{V}}\nolimits(\Theta_{0}) \ge \mathop{\mathbf{J}}%
\nolimits^{-1}(\Theta_{0})=\frac{1}{t\tilde{\lambda}}\bar{\mathop{\mathbf{J}}
\nolimits}^{-1}(\Theta_{0}),  \label{eq:CRbound3}
\end{equation}
that is, in the the asymptotic regime, the errors (the covariance) of the
maximum likelihood estimates should be decreasing as inversely proportional
to the data acquisition time $t$. Consequently, from Eqs.~(\ref{eq:Bures})
and (\ref{eq:limitTM}), we have 
\begin{equation}
2(1-F(\rho _{\Theta _{0}},\rho _{\hat{\Theta}})) \ge \frac{1}{4t\tilde{%
\lambda}}\mathop{\mathbf{Tr}}\nolimits[\mathop{\mathbf{J}}%
\nolimits^{SLD}(\Theta _{0})\,\bar{\mathop{\mathbf{J}}\nolimits}%
^{-1}(\Theta_{0})],  \label{eq:limitTM2}
\end{equation}
or, equivalently, 
\begin{eqnarray}
&\,& \ln [1-F(\rho _{\Theta _{0}},\rho _{\hat{\Theta}})]  \nonumber \\
&\ge& -\ln [\lambda]+\ln [\frac{1}{8}\mathop{\mathbf{Tr}}\nolimits[%
\mathop{\mathbf{J}}\nolimits^{SLD}(\Theta _{0})\,\bar{\mathop{\mathbf{J}}%
\nolimits}^{-1}(\Theta_{0})]].  \label{eq:loglimitTM}
\end{eqnarray}
The logarithm of the average Bures distance between the \textit{true} state
and the estimated state is thus supposed to be decreasing proportional to
the logarithm of the nuisance parameter, $\lambda=t\tilde{\lambda}$ (the
first term in the right-hand side of Eq.~(\ref{eq:loglimitTM})), and all
state-dependent properties appear as the intercept on the axis of ordinates
(the second term in the right-hand side of Eq.~(\ref{eq:loglimitTM})).

\begin{figure}[tbp]
\includegraphics[width=0.95\linewidth]{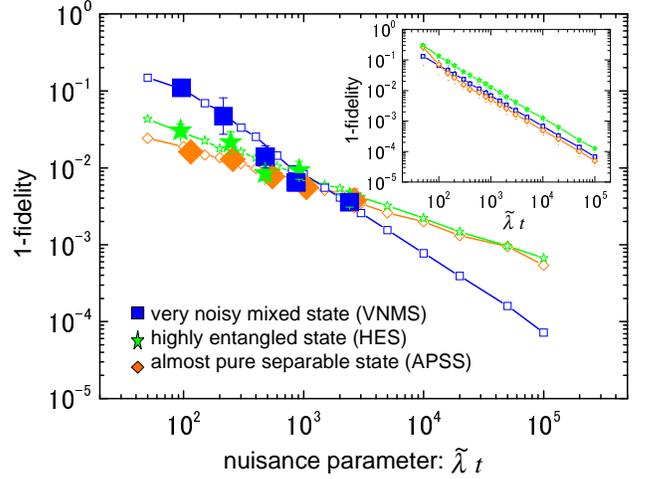}
\caption{The average Bures distances (divided by 2) between the \textit{true}
states and the states estimated by the MLE are shown as a function of the
nuisanece parameter $\tilde{\protect\lambda}\,t$. The filled plots represent
experimental results (the vertical error bars correspond to one standard
deviation) and the blank plots represent numerical simulations. The inset
shows their asymptotic lower bounds.}
\label{fig:mle}
\end{figure}

We note that there are some difficulties for practically calculating the
asymptotic lower bound. First, as can be seen in Eq.~(\ref{eq:Fisher}), the
calculation of the Fisher information matrix $J_{ij}(\Theta _{0})$ in Eq.~(%
\ref{eq:limitTM}) includes 16 infinite sum for computing an average over the
probability mass function (\ref{eq:pdf}). For circumventing this difficulty,
we make an approximation; 
\begin{equation}
J_{ij}(\Theta _{0})\approx \frac{1}{1000}\sum_{t=1}^{1000}(\frac{\partial }{%
\partial \theta ^{i}}\ln [P(N_{0}^{(t)}|\Theta )]\frac{\partial }{\partial
\theta ^{j}}\ln [P(N_{0}^{(t)}|\Theta ))],  \label{eq:AvFisher}
\end{equation}
where $\{N_{0}^{(t)}\}$ is numerically simulated (1000 times) according to
the Gaussian approximation of $P(N|\Theta _{0})$ in Eq.~(\ref{eq:pdf}).
Second, the inverse of the Fisher information matrix (\ref{eq:AvFisher})
should be derived from the so-called \textit{Moore-Penrose generalized
inverse} \cite{MPinverse} in case the determinant of the Fisher information
matrix becomes zero (the Moore-Penrose generalized inverse provides the
unique and well-behaved inverse even for such degenerate matrices). Third,
the SLD $\mathop{\mathbf{L}}\nolimits_{i}^{S}(\Theta )$ of Eq.~(\ref%
{eq:SLDFisher}) is not uniquely determined from Eq.~(\ref{eq:SLD}), except
for \textit{non-degenerate} states (the states whose eigenvalues are all
non-zero values, that is, the strictly positive states, or the rank-4 states
for our specific example). This problem was investigated by Fujiwara,
Nagaoka \cite{FN1995}, Hayashi \cite{Hayashi1998}, and Matsumoto \cite%
{Matsumoto2002} for pure states, and Fujiwara \cite{FujiwaraPhD}, Matsumoto %
\cite{MatsumotoPhD}, and Fujiwara, Nagaoka \cite{FN1999} for more general
degenerate states. According to their results, any SLDs derived from Eq.~(%
\ref{eq:SLD}) results in the same SLD Fisher information matrix $%
\mathop{\mathbf{J}}\nolimits^{SLD}(\Theta _{0})$. For this reason, we solved
Eq.~(\ref{eq:SLD}) by using the Moore-Penrose generalized inverse, which
provides an unique solution for $\mathop{\mathbf{L}}\nolimits_{i}^{S}(\Theta
)$ and regarded it as a representative of the SLDs.

To see whether the theoretical predictions, Eq.~(\ref{eq:loglimitTM}), are
truly observed in our experiments, the average Bures distances between the 
\textit{true} states and the estimated states as a function of ensemble
size, $\tilde{\lambda}\,t$, (i.e., the nuisance parameter) are presented in
Fig.~\ref{fig:mle}, where the both axes of Fig.~\ref{fig:result} are
converted into their logarithms. The calculated asymptotic lower bounds of
the average Bures distances are also shown in the inset of Fig.~\ref{fig:mle}%
. The slight deviation from the linear slope in the small nuisance parameter
regime in the inset of Fig.~\ref{fig:mle}, is probably due to the mismatch
between the Gaussian approximation used for producing simulated data $%
\{N_{0}^{(t)}\}$ and their genuine distribution, i.e., Poisson distribution.

In Fig.~\ref{fig:mle}, the average Bures distances clearly depend on what
kind of states are to be estimated. However, there are discrepancies between
the asymptotic lower bounds (the inset of Fig.~(\ref{fig:mle})) and the
experimental results. The discrepancies in the small ensemble region (the
left-hand side of Fig.~(\ref{fig:mle})) might be explained by higher order
effect of the errors \cite{BH1996} (i.e., by the deviations from the first
order approximation of the Bures distance in Eq.~(\ref{eq:Riemann})). On the
other hand, the discrepancies in the large ensemble region (the right-hand
side of Fig.~(\ref{fig:mle})) cannot be explained by such higher order
effects. Except for the results of the VNMS, the expected asymptotic
behavior (i.e., the inverse proportionality to the ensemble size) is not
observed, even in the results of the numerical simulations. On the other
hand, the simulations were carried out including no systematic errors. These
facts negate a possibility that the cause of the above discrepancies could
stem from perturbations of the experimental condition. Another possibility
is that the \textit{true} state may be slightly \textit{biased} due to the
fact that we determined it by the estimation, thus the (possibly biased) 
\textit{true} state plays a major role in the discrepancy. However, for the
numerical simulations, this \textit{true} state is a bona-fide state, which
is used as the source to produce the artificial-coincidence-count. This fact
nullifies the latter possibility, too.

In the next section, we will elaborate on a possible reason for the
discrepancies, and introduce a new estimation strategy based on the \textit{%
Akaike's information criterion} \cite{Akaike1974} for reducing the
discrepancies and approaching the asymptotic lower bound, Eq.~(\ref%
{eq:loglimitTM}).


\section{Akaike's information criterion}

\label{sec:AIC}

Remember that we implicitly assumed that the parametric model of the quantum
states (given by Eq.~(\ref{eq:DM2}) and (\ref{eq:Cholesky})) is full-rank,
that is, we parametrized the quantum states with 16 parameters (including
the nuisance parameter); see Appendix \ref{app:tomography} and \ref{app:MLE}%
. However degenerate states, such as the APSS or the HES, might be
completely characterized by less than 16 parameters. Subsequently, the
surplus parameters give rise to an ambiguity in the numerical procedure of
the MLE, i.e., in finding the maximum of the likelihood function, (\ref%
{eq:pdf}). These procedures were executed by \texttt{FindMinimum}, a
function of MATHEMATICA 4.0, which is employing the multi-dimensional Powell
algorithm, as in Ref.~\cite{JKMW2001}. However, the minimum found by this
function is not necessarily the \textit{global} minimum. Note that there are
some ways to circumvent the problem of such \textit{local} minimums of
likelihood function, e.g., by using the quasi-Newton methods or employing a
sophisticated iterative procedures, \textit{the expectation-maximization
(EM) algorithm followed by unitary transformation}, which is due to \u{R}eh%
\'{a}\u{c}ek \textit{et al.} \cite{RHJ2001}. Here we will, however, give a
rather simple but thought-provoking procedure based on the so-called \textit{%
Akaike's information criterion (AIC)} \cite{Akaike1974}, which eliminates
the redundant parameters.

The AIC is defined by 
\begin{equation}
AIC^{(k)}(\Theta )=-2\,\ln [P^{(k)}(N|\Theta )]+2\,k,  \label{eq:AIC}
\end{equation}
where $k$ is the number of independent parameters and $\ln
[P^{(k)}(N|\Theta)]$ is the log-likelihood function for the quantum state $%
\rho _{\Theta }^{(k)}$ which parametrized by $k$ parameters \cite{note_AIC}.
When there are several hypothetical models (with different number of
parameters) for estimating a certain state, the model which attains the
smallest AIC can be regarded as the most appropriate model because of the
following justification. In Appendix~\ref{app:MLE}, for explaining the MLE,
we used the fact that the approximation~(\ref{eq:Mloglikelihood}) is valid
in the asymptotic region. What Akaike found \cite{Akaike1974} is that there
is a difference between the mean of the maximum log-likelihood function
(right-hand side of (\ref{eq:Mloglikelihood})) and the maximum
log-likelihood function derived by the obtained data (left-hand side of (\ref%
{eq:Mloglikelihood})), and the difference can be approximately given by 
\begin{eqnarray}
\Delta ^{(k)}(\hat{\Theta}) &=&\frac{1}{t}\ln [P^{(k)}(N|\hat{\Theta})]-%
\mathop{\mathbf{E}}\nolimits_{\Theta _{0}} [\ln [P^{(k)}(N|\hat{\Theta}%
)]|_{t=1}]  \nonumber \\
&\approx &\frac{k}{t}.  \label{eq:difference}
\end{eqnarray}
Taking this correction into account, the Kullback-Leibler distance between
the \textit{true} probability mass function $P_{0}(N)$ and its parametric
model $P(N|\Theta)$, i.e., Eq.~(\ref{eq:relativeE}), can be minimized by
reducing the value, 
\begin{eqnarray}
-\mathop{\mathbf{E}}\nolimits_{\Theta _{0}} [\ln [P^{(k)}(N|\hat{\Theta}%
)]|_{t=1}] &=&-\frac{1}{t}\ln [P^{(k)}(N|\hat{\Theta})]+\frac{k}{t} 
\nonumber \\
&=&\frac{1}{2\,t}AIC^{(k)}(\hat{\Theta}),  \label{eq:AIC2}
\end{eqnarray}
with respect to the estimators $\{\hat{\Theta}\}$. Therefore, if we choose
the model which minimizes the AIC (\ref{eq:AIC}) among several alternative
parametric models, it is ensured that this model is the closest to the 
\textit{true} one from the viewpoint of the Kullback-Leibler distance. The
resultant estimate is called \textit{minimum AIC estimate (MAICE)} \cite%
{Akaike1974}. When a maximum likelihood estimates of a certain model is
almost identical to that of another model, the MAICE becomes the one defined
with the smaller number of the parameters. The definition of the MAICE gives
the mathematical formulation of the principle of parsimony in model
selection.

The importance of this new strategy might be more noticeable in estimating
quantum states in the infinite-dimensional Hilbert space, e.g., in
estimating Wigner function~\cite{SBRF1993} or density matrix in the number
state representation~\cite{SBPMM1996}. In this situation, somehow vague
Fourier-frequency cutoff or truncation of an infinite-dimensional density
matrix to finite-dimensional density matrix is introduced in executing the 
\textit{inverse Radon transformation} or \textit{quantum-state sampling},
respectively~\cite{Leonhardt1997}. We note that Gill and Gu\c{t}\u{a},
recently, made a first attempt addressing this issue \cite{GG2003}.

Specifically, for estimating the two qubits, we use 
\begin{equation}
\rho_{\Theta}^{(k)}=\frac{T_{\Theta}^{(k)}(T_{\Theta}^{(k)})^{\dagger}}{%
\mathop{\mathbf{Tr}}\nolimits[T_{\Theta}^{(k)}(T_{\Theta}^{(k)})^{\dagger}]},
\label{eq:DM4}
\end{equation}
where 
\begin{eqnarray}
T_{\Theta}^{(16)} &=& \left[ 
\begin{array}{cccc}
\theta^1 & 0 & 0 & 0 \\ 
\theta^2+i\theta^3 & \theta^8 & 0 & 0 \\ 
\theta^4+i\theta^5 & \theta^9+i\theta^{10} & \theta^{13} & 0 \\ 
\theta^6+i\theta^7 & \theta^{11}+i\theta^{12} & \theta^{14}+i\theta^{15} & 
\theta^{16}%
\end{array}
\right],  \nonumber \\
T_{\Theta}^{(15)} &=& \left[ 
\begin{array}{cccc}
\theta^1 & 0 & 0 & 0 \\ 
\theta^2+i\theta^3 & \theta^8 & 0 & 0 \\ 
\theta^4+i\theta^5 & \theta^9+i\theta^{10} & \theta^{13} & 0 \\ 
\theta^6+i\theta^7 & \theta^{11}+i\theta^{12} & \theta^{14}+i\theta^{15} & 0%
\end{array}
\right],  \nonumber \\
T_{\Theta}^{(12)} &=& \left[ 
\begin{array}{cccc}
\theta^1 & 0 & 0 & 0 \\ 
\theta^2+i\theta^3 & \theta^8 & 0 & 0 \\ 
\theta^4+i\theta^5 & \theta^9+i\theta^{10} & 0 & 0 \\ 
\theta^6+i\theta^7 & \theta^{11}+i\theta^{12} & 0 & 0%
\end{array}
\right],  \nonumber \\
T_{\Theta}^{(7)} &=& \left[ 
\begin{array}{cccc}
\theta^1 & 0 & 0 & 0 \\ 
\theta^2+i\theta^3 & 0 & 0 & 0 \\ 
\theta^4+i\theta^5 & 0 & 0 & 0 \\ 
\theta^6+i\theta^7 & 0 & 0 & 0%
\end{array}
\right],
\end{eqnarray}
thus, $\rho_{\Theta}^{(16)}$, $\rho_{\Theta}^{(15)}$, $\rho_{\Theta}^{(12)}$%
, and $\rho_{\Theta}^{(7)}$ are representing the rank-4, rank-3, rank-2, and
rank-1 density matrices, respectively. Then the AICs are respectively given
by 
\begin{eqnarray}
AIC^{(16)}(\hat{\Theta}) &=& -2\,\ln[P^{(16)}(N|\hat{\Theta})] + 2 \times 16,
\nonumber \\
AIC^{(15)}(\hat{\Theta}) &=& -2\,\ln[P^{(15)}(N|\hat{\Theta})] + 2 \times 15,
\nonumber \\
AIC^{(12)}(\hat{\Theta}) &=& -2\,\ln[P^{(12)}(N|\hat{\Theta})] + 2 \times 12,
\nonumber \\
AIC^{(7)}(\hat{\Theta}) &=& -2\,\ln[P^{(7)}(N|\hat{\Theta})] + 2 \times 7,
\label{eq:AIC3}
\end{eqnarray}
where $P^{(k)}(N|\hat{\Theta})$ is the same form of Eq.~(\ref{eq:pdf}) but
replacing $M_{\nu}(\Theta)$ with 
\begin{equation}
M^{(k)}_{\nu}(\Theta) = \mathop{\mathbf{Tr}}\nolimits[|m_{\nu}\rangle
\langle m_{\nu}|T_{\Theta}^{(k)}(T_{\Theta}^{(k)})^{\dagger}].
\label{eq:meanPro_k}
\end{equation}
Among these models, we can choose the one which minimizes the AIC. As an
example, for one of the typical experimental data of coincidence counts for
the VNMS (data acquisition time: 5s), $\{N\}$=\{615, 553, 613, 605, 550,
576, 596, 609, 575, 622, 577, 601, 574, 569, 591, 569\}, we have the
following AICs; $AIC^{(16)}=163.4$, $AIC^{(15)}=201.3$, $AIC^{(12)}=349.9$
and $AIC^{(7)}=2899.3$. Therefore we choose the rank-4 model $%
\rho_{\Theta}^{(16)} $. On the other hand, for one of the typical data for
the APSS (data acquisition time: 5s), $\{N\}$=\{42, 45, 25, 2504, 60, 56,
31, 33, 1309, 1431, 1148, 1125, 514, 487, 576, 599\}, we have; $%
AIC^{(16)}=152.8$, $AIC^{(15)}=150.8$, $AIC^{(12)}=146.3$ and $%
AIC^{(7)}=208.9$. Thus the rank-2 model $\rho_{\Theta}^{(12)}$ is chosen.

It is possible to think about the other hypothetical models, e.g., separable
model (which has 7 parameters), or separable and also rank-1 model (which
has 5 parameters), but for simplicity, our analyses were confined to the
above 4 models.

\begin{figure}[tbp]
\includegraphics[width=0.95\linewidth]{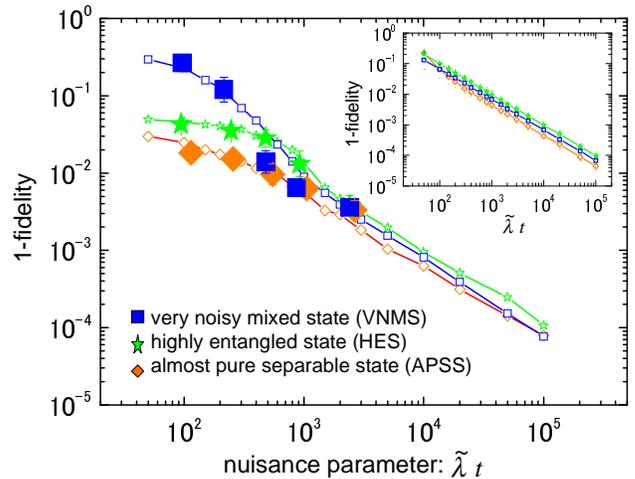}
\caption{The average Bures distances (divided by 2) between the \textit{true}
states and the MAICEs of the states are shown as a function of the nuisanece
parameter $\tilde{\protect\lambda}\,t$. The filled plots represent
experimental results (the vertical error bars correspond to one standard
deviation) and the blank plots represent numerical simulations. The inset
shows their asymptotic lower bounds.}
\label{fig:aic}
\end{figure}

Figure~\ref{fig:aic} shows the average Bures distances between the \textit{%
true} states and the estimated states obtained by employing the new
estimation strategy. Their asymptotic lower bounds are also exhibited in the
inset of Fig.~\ref{fig:aic}. These asymptotic values were calculated
according to Eq.~(\ref{eq:limitTM}). Note that since the \textit{true} state
here was also determined by the MAICE instead of MLE, \textit{true}
state for the APSS and that for the HES resulted in rank-2 density matrices.

As a result of the reduction of the parameters, the MAICEs substantially
reduces the discrepancies between the asymptotic lower bounds (the inset)
and the experimental results (filled plots: experiments, blank plots: Monte
Carlo simulations) comparing to the previous results (Fig.~\ref{fig:mle}).
Moreover, in the region where the data acquisition time $t$ greater than 2s,
i.e., $\lambda =t\tilde{\lambda}\geq 10^{3}$, the lower bounds of Eq.~(\ref%
{eq:limitTM2}) are almost achieved. This is the case even for estimating
degenerate states such as the APSS and the HES. Note also that the intercepts of the asymptotic values on the axis of ordinates shown in the inset of Fig.~%
\ref{fig:aic} are slightly lowered comparing to the previous ones (the inset
of Fig.~\ref{fig:mle}).

Here we remark that while the numerical simulations continue up to $\lambda
=100000$, the maximum data acquisition time of each experiment is 5s, which
corresponds to $\lambda \approx 2500$. This is because the systematic errors
mentioned above might be getting significant around $\lambda \approx 10000$.

The decreasing rate of the Bures distance for the APSS deviates slightly
from the ideal value, -1. This might be due to the residue of the redundant
parameters even after making model-inquiry among the above 4 models, because
for APSS, the another model (e.g., separable model or separable and rank-1
model) might be more suitable. Thus the further reduction of the parameters
might be possible. The discrepancies in small ensemble region (left-hand
side of Fig.~(\ref{fig:aic})) may be explained by higher order effect of the
errors \cite{BH1996} as is mentioned in the previous section.

What kind of factor does dominantly affect the accuracy of the estimation?
This question has not been perfectly answered so far. In the general setting
of the quantum parameter estimation problem \cite%
{Helstrom1976,Holevo1982,YL1973}, any kinds of measurements represented as
the \textit{positive operator-valued measures (POVMs)} \cite{NCtext2000} are
allowed to be utilized. Then not only inseparable projective measurements on
the two qubits but even \textit{collective} measurements on whole ensembles
are allowed. In this setting, it has been known that the \textit{%
non-commutativity} of quantum mechanics has significant influence on the
attainable lower bounds on errors in estimating quantum states with multiple-parameter \cite{MatsumotoPhD, FN1999}. Although significant
progress has been made \cite{Holevo2001,ANtext2000,FujiwaraPhD,MatsumotoPhD,Hayashi1998,FN1999,GM2000,Matsumoto2002}, finding the asymptotically optimal measurement strategy and obtaining the achievable lower bounds on the errors in estimating quantum states with multiple-parameter are still important open problems.

On the other hand, in our setting, the measurement strategy we employed is
not such a optimal \textit{collective}-measurement strategy, but \textit{%
local} tomographic measurements represented by (\ref{eq:projectors}).
Nonetheless, our results reveal another aspects of the quantum state
estimation, that is, the nature of local measurements. Figure~\ref{fig:aic}
shows that the errors in estimating the entangled state, i.e., the HES
(which has small entropy but large entanglement; see Fig.~\ref{fig:devmatrix}%
) is the largest among the three states in the asymptotic region. Thus, the
existence of entanglement seems to degrade the accuracy of the estimation if
the measurements are performed locally.


\section{alternative measurement strategy}

\label{sec:collective}

In this section, we discuss the alternative measurement strategy for two
qubits. Since it may be extremely difficult to experimentally realize
optimal \textit{collective} measurements, the following discussion is
restricted to projective measurements on just one sample in the ensemble,
i.e., on two qubits. Note that there is another favorable measurement
strategy, that is, \textit{self-learning measurements}~\cite%
{FKF2000,HRBNTW2002}. However, to the best of our knowledge, the lower bound
on errors in estimating with this type of adaptive-measurement strategy is
still missing.

It is reasonable to expect that if we employ measurements on the \textit{%
inseparable} projectors on two qubits, the errors in estimating the
entangled states might be reduced. For inspecting whether this expectation
is true or not, the following specific projective measurements: 
\begin{equation}
\begin{array}{c}
\frac{1}{2}\,(|HH\rangle +|VV\rangle )(\langle HH|+\langle VV|) \\ 
\frac{1}{2}\,(|HH\rangle -|VV\rangle )(\langle HH|-\langle VV|) \\ 
\frac{1}{2}\,(|HV\rangle +|VH\rangle )(\langle HV|+\langle VH|) \\ 
\frac{1}{2}\,(|HV\rangle -|VH\rangle )(\langle HV|-\langle VH|) \\ 
\frac{1}{2}\,(|HD\rangle +|VX\rangle )(\langle HD|+\langle VX|) \\ 
\frac{1}{2}\,(|HD\rangle -|VX\rangle )(\langle HD|-\langle VX|) \\ 
\frac{1}{2}\,(|HX\rangle +|VD\rangle )(\langle HX|+\langle VD|) \\ 
\frac{1}{2}\,(|HR\rangle +|VL\rangle )(\langle HR|+\langle VL|) \\ 
\frac{1}{2}\,(|HR\rangle -|VL\rangle )(\langle HR|-\langle VL|) \\ 
\frac{1}{2}\,(|HL\rangle +|VR\rangle )(\langle HL|+\langle VR|) \\ 
|H\rangle \langle H|\otimes \frac{1}{2}I \\ 
\frac{1}{2}I\otimes |H\rangle \langle H| \\ 
|D\rangle \langle D|\otimes \frac{1}{2}I \\ 
\frac{1}{2}I\otimes |D\rangle \langle D| \\ 
|R\rangle \langle R|\otimes \frac{1}{2}I \\ 
\frac{1}{2}I\otimes |R\rangle \langle R|,%
\end{array}
\label{eq:projectors2}
\end{equation}
are employed as an alternative to the \textit{local} tomographic
measurements (\ref{eq:projectors}). Here, as mentioned in Appendix \ref%
{app:tomography}, $|D \rangle=\frac{1}{\sqrt{2}}(|H \rangle+|V \rangle)$, $%
|X \rangle =\frac{1}{\sqrt{2}}(|H \rangle-|V \rangle)$, $|R \rangle =\frac{1%
}{\sqrt{2}}(|H \rangle+i|V \rangle)$, and $|R \rangle =\frac{1}{\sqrt{2}}%
(|H\rangle-i|V\rangle)$, as $|H \rangle$ and $|V \rangle$ being the
horizontal polarization state and vertical one, respectively. This set of 16
projective measurements includes 10 inseparable projectors, and satisfies
the condition of tomographic measurements, which is presented in Appendix~%
\ref{app:tomography}.

\begin{figure}[tbp]
\includegraphics[width=\linewidth]{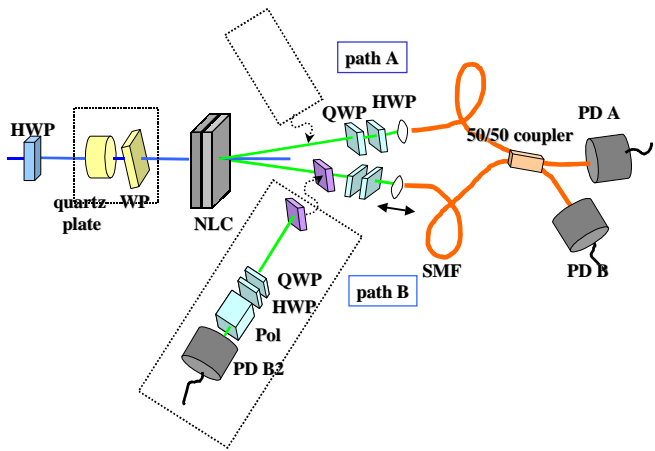}
\caption{Outline of the proposed experimental setup for realizing the
projenctive measurements (\ref{eq:projectors2}). See text for details.}
\label{fig:BSA}
\end{figure}

These projective measurements can be realized by slightly modifying the
interferometric Bell-state analyzer \cite{MMWZ1996}. Figure~\ref{fig:BSA}
shows the proposed experimental setup for realizing the projective
measurements (\ref{eq:projectors2}). For 10 inseparable-projective
measurements in (\ref{eq:projectors2}), the down-converted photons are
coupled into single mode optical fibers (SMFs) and mixed at 50/50 coupler.
Then if the optical path length of two photons are appropriately adjusted
and the effects of birefringence in the SMFs are compensated by fiber
polarization controller, only two photon whose state of polarization belongs
to anti-symmetric subspace, (singlet subspace) contribute the coincidence
counts of photon detector (PD) A and B. This coincidence measurement is
equivalent to one of the inseparable projective measurements, $\frac{1}{2}%
\,(|HV\rangle -|VH\rangle )(\langle HV|-\langle VH|)$. The other 9
inseparable measurements can be straightforwardly realized by the local
unitary transformation of the state of polarization using half-wave plates
(HWP) and quarter-wave plates (QWP) before coupling photons into the SMFs.

For remained 6 local-projective measurements in (\ref{eq:projectors2}), the
state of one-photon polarization is projected on the particular state, e.g., 
$|H\rangle \langle H|$, by inserting a mirror into the path B (path A) and
using HWP, QWP, polarizer, and PD B2 (PD A2 (not shown)) as shown in the
dotted box in Fig.~\ref{fig:BSA}. Another photon is propagated to either PD
A or B. The coincidence measurements of PD B2 (PD A2) and either PD A or PD
B are served as the local-projective measurements, e.g., $\frac{1}{2}%
I\otimes |H\rangle \langle H|$ ($|H\rangle \langle H|\otimes \frac{1}{2}I$).

For utilizing this strategy as an alternative to the local one (\ref%
{eq:projectors}), it is vital to minimize the systematic errors due to
imperfect intensity interference in the inseparable measurements. In order
to achieve required high visibility of interference, the distinguishability
in any degree of freedom of two photons other than the polarization should
be reduced. By using SMFs for enhancing spacial-mode overlap of two photons,
we expect that such systematic errors due to the spacial degree of freedom
might be reduced to some extent. In the recent experiments, the visibilities
of this interference exceeding 98\% \cite{HDFF2002} and even reaching 99.4\% %
\cite{PF2003} were reported.

\begin{figure*}[tbp]
\includegraphics[width=\linewidth]{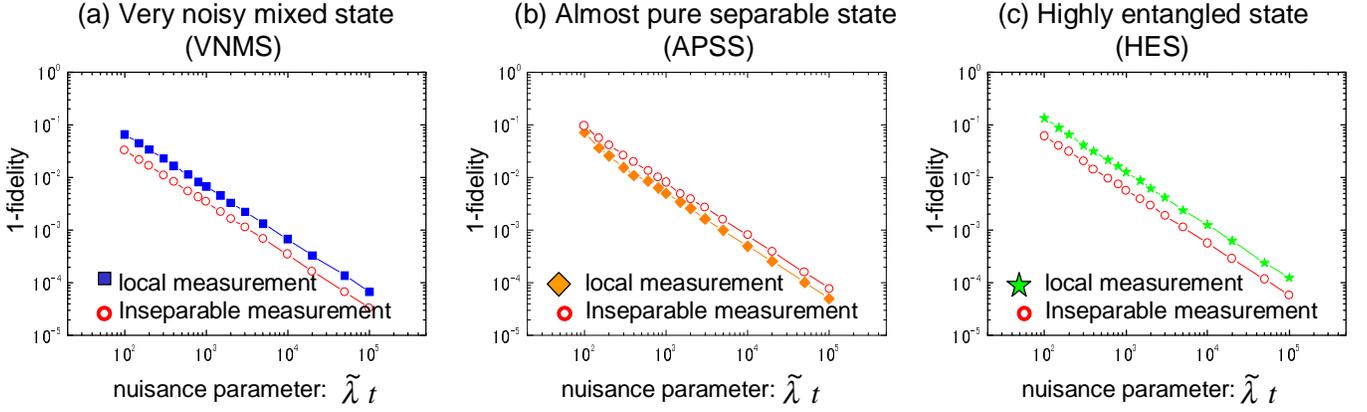}
\caption{The comparison between the asymptotic lower bounds Eq.~(\ref%
{eq:loglimitTM}) for \textit{inseparable} tomographic measurements (\ref%
{eq:projectors2}) and for the conventional \textit{local} ones (\ref%
{eq:projectors}).}
\label{fig:ubasis}
\end{figure*}

The comparison between the asymptotic lower bounds Eq.~(\ref{eq:loglimitTM})
for the above \textit{inseparable} measurements (\ref{eq:projectors2}) and
the conventional \textit{local} ones (\ref{eq:projectors}) is presented in
Fig.~\ref{fig:ubasis}. As expected, the improvement of the accuracy in
estimating the HES can be found in Fig.~\ref{fig:ubasis} (c), although the
accuracy is decreased in estimating the APSS as can be seen in Fig.~\ref%
{fig:ubasis} (b). As indicated in Fig.~\ref{fig:ubasis} (a), even for the
VNMS, which has no entanglement at all (see Fig.~\ref{fig:devmatrix}), the
inseparable measurements (\ref{eq:projectors2}) are working better than the
separable ones (\ref{eq:projectors}). This rather surprising result might be
viewed as the \textit{non-locality without entanglement} in quantum state
estimation \cite{MP1995,Hayashi1998,GM2000}. We conjecture that for the
mixed states like the VNMS, no \textit{local} tomographic measurements can
attain the same accuracy achieved by the inseparable measurements presented
in (\ref{eq:projectors2}). This phenomenon may stem from the fact that the
mixed states can be represented as the classical mixture of the entangled
states as well as that of the product states.


\section{conclusion}

\label{sec:conclusion} We presented quantitative analysis concerning the
accuracy of the quantum state estimation, and demonstrated that they depend
both on the states to be estimated and on the measurement strategies. For
this purpose, the SPDC process was employed for experimentally preparing
various ensembles of the bi-photon polarization states and the AIC based new
estimation strategy, i.e., the MAICE was introduced for eliminating the
numerical problems in the estimation procedures. Our results showed errors
of the estimated density matrices decreased as inversely proportional to the
ensemble size for all of the three states we examined (the VNMS, the APSS,
and the HES) in the asymptotic region. Besides, it was revealed that the
existence of entanglement degrade the accuracy of the estimation when the
measurements were performed locally on two qubits. The performance of the
alternative measurement strategy, which included the projective measurements
on inseparable bases, was numerically examined, and we found that the
inseparable measurements improved the accuracy in estimating the VNMS as
well as the HES.

Further study of the quantum state estimation is sure to pave the way for
understanding the ultimate rule for acquiring information from quantum
systems.


\begin{acknowledgments}
We are grateful to Tohya~Hiroshima, Satoshi~Ishizaka, Bao-Sen~Shi,
Akihisa~Tomita, Masahito~Hayashi, Masahide~Sasaki, and Prof.~Osamu~Hirota
for valuable discussions and encouragements, to Shunsuke~Kono and
Kenji~Kazui for their technical support, and to Kwangseuk~Kyhm for reviewing
the manuscript. We also would like to thank Jaroslav \u{R}eh\'a\u{c}ek for
useful correspondence.
\end{acknowledgments}

\appendix

\section{}
In Apps.~\ref{app:tomography} and \ref{app:MLE}, we give a brief review of
tomographic measurements and maximum likelihood estimation (MLE),
respectively, in accordance with Ref.~\cite{JKMW2001}. Readers who are
familiar with two issues can skip these two Apps. We mention the
Cram\'er-Rao inequality and the Fisher information matrix for providing the
optimality of the MLE in Apps.~\ref{app:CR bound}.


\subsection{\label{app:tomography}tomographic measurements}

With the standard Pauli matrices $\{\sigma _{i}\}_{i=1}^{3}$ supplemented
with the identity matrix $\sigma _{0}=I$, an arbitrary density matrix of two
qubits can be represented in Hilbert-Schmidt space as a parametric
statistical model: 
\begin{equation}
\rho _{\Phi }=\sum_{i=0}^{3}\sum_{j=0}^{3}\frac{1}{4}\ (\sigma _{i}
\otimes\sigma _{j})\ \phi ^{i,j}=\sum_{\mu =0}^{15}\Gamma _{\!\mu }\ \phi
^{\mu },  \label{eq:DM}
\end{equation}
where $\Gamma _{\!4i+j}=\frac{1}{4}\ (\sigma _{i}\otimes \sigma _{j})$ and $%
\phi ^{4i+j}=\phi ^{i,j}$. Here $\{\phi ^{\mu }\}_{\mu =0}^{15}$ are assumed
to be real. From the trace condition of a density matrix, $\phi ^{0}$ is
equal to one. Note that the above parametric model in Hilbert-Schmidt space
does not ensure positivity condition of a density matrix, the problem of
positivity is revisited in \ref{app:MLE}.

When we try to estimate quantum states as the parametric statistical model
of Eq.~(\ref{eq:DM}), we should perform some kinds of measurements. Suppose
that the measurements are represented by projectors $|m_{\nu}\rangle \langle
m_{\nu}|$. Imagining coincidence counting measurements on bi-photon
polarization states as a concrete example, the projectors correspond to a
certain polarization states. After carrying out the measurements for data
acquisition time $t$, the results are given by 
\begin{equation}
n_{\nu}=\lambda \mathop{\mathbf{Tr}}\nolimits[|m_{\nu}\rangle \langle
m_{\nu}|\rho _{\Phi }],  \label{eq:coincidence}
\end{equation}
where 
\begin{equation}
\lambda =\tilde{\lambda}\,t,  \label{eq:nuisance}
\end{equation}
is the coincidence counts without polarizers for the data acquisition time $%
t $. Then, $\tilde{\lambda}$ is the coincidence counting rate. Although our
attention is focused on the 15 parameters $\{\phi^{\mu}\}_{\mu =0}^{15}$, $%
\tilde{\lambda}$ is also \textit{a priori} unknown. Therefore, $\tilde{%
\lambda}$ is appended to the list of the parameters for estimating the
states. The parameter $\lambda$ is thus called the \textit{nuisance parameter%
}. Using Eq.~(\ref{eq:DM}), Eq.~(\ref{eq:coincidence}) becomes 
\begin{equation}
n_{\nu}=\lambda \sum_{\mu =0}^{15}\langle m_{\nu}|\Gamma _{\!\mu }\
|m_{\nu}\rangle \phi ^{\mu }=\lambda \sum_{\mu =0}^{15}B_{u \mu }\phi^{\mu }.
\label{eq:coincidence2}
\end{equation}
Eq.~(\ref{eq:coincidence2}) provides a linear relationship between the 16
parameters $\{\phi ^{\mu }\}_{\mu =1}^{15}$ and $\lambda $, and the
measurement results $\{n_{\nu}\}$. Subsequently, we can derive a necessary
and sufficient condition of the measurement for determining these
parameters, that is, the matrix $B_{u \mu }$ has an inverse (thus the
measurement should consist of at least 16 projectors). Measurements that
satisfy the above condition are called \textit{tomographic measurements} %
\cite{JKMW2001}.

A specific instance of tomographic measurements $\{|m_{u} \rangle\langle
m_{u}|\}_{u=1}^{16}$ are \cite{note_TM}: 
\begin{equation}
\begin{array}{cccc}
|HH \rangle\langle HH| & |HV \rangle\langle HV| & |HD \rangle\langle HD| & 
|HL \rangle\langle HL| \\ 
|VH \rangle\langle VH| & |VV \rangle\langle VV| & |VD \rangle\langle VD| & 
|VL \rangle\langle VL| \\ 
|DH \rangle\langle DH| & |DV \rangle\langle DV| & |DD \rangle\langle DD| & 
|DL \rangle\langle DL| \\ 
|RH \rangle\langle RH| & |RV \rangle\langle RV| & |RD \rangle\langle RD| & 
|RL \rangle\langle RL|, \label{eq:projectors}%
\end{array}%
\end{equation}
where $|D \rangle=\frac{1}{\sqrt{2}}(|H \rangle+|V \rangle)$, $|X \rangle =%
\frac{1}{\sqrt{2}}(|H \rangle-|V \rangle)$, $|R \rangle =\frac{1}{\sqrt{2}}%
(|H \rangle+i|V \rangle)$, and $|R \rangle =\frac{1}{\sqrt{2}}(|H\rangle-i|V
\rangle)$, as $|H \rangle$ and $|V \rangle$ being the horizontal
polarization state and vertical one, respectively. Here $|HH \rangle\langle
HH|$ means $(|H \rangle \otimes |H \rangle )(\langle H|\otimes \langle H|)$.

From the measurement results $\{n_{\nu}\}_{\nu =1}^{16} \equiv \{N\}$, we
can solve linear equation, Eq.~(\ref{eq:coincidence2}), with respect to the
parameters, $\{\phi ^{\mu }\}_{\mu =1}^{15}$ and $\lambda$. As a result, the
quantum state of the form Eq.~(\ref{eq:DM}) can be uniquely reconstructed.
The solutions are explicitly expressed as 
\begin{equation}
\phi ^{\mu }=\frac{1}{\lambda }\sum_{\nu =1}^{16}[B_{u \mu }]^{-1}n_{\nu}.
\label{eq:linear tomography}
\end{equation}
This estimation strategy is called the \textit{linear tomography} \cite%
{JKMW2001}.


\subsection{\label{app:MLE} Maximum likelihood estimation}

The flaw of the linear tomography in Sec.~\ref{app:tomography} is two-fold.
One is that there are no considerations about its optimality, another is
that the parametric model for linear tomography, Eq.~(\ref{eq:DM}), does not
ensure the positivity condition of the density matrix as mentioned before.
The solution for these flaws is to use \textit{maximum likelihood estimation
(MLE)} \cite{JKMW2001,BDPS1999,RHJ2001}.

Density matrix, which satisfy the positive condition and also the trace
condition, can be written as 
\begin{equation}
\rho _{\Theta}=\frac{T_{\Theta}T_{\Theta}{}^{\dagger}}{\mathop{\mathbf{Tr}}%
\nolimits[T_{\Theta}T_{\Theta}{}^{\dagger}]},  \label{eq:DM2}
\end{equation}
where $T_{\Theta}$ is assumed to be a normal matrix. Then, following Ref.~%
\cite{JKMW2001,BDPS1999}, we adopt the complex lower triangular matrix
parametrized by 16 real parameters, $\{\theta^{\mu}\}_{\mu =1}^{16} \equiv
\{\Theta\}$, (Cholesky decomposition) as the normal matrix $T_{\Theta}$. It
is explicitly written as 
\begin{equation}
T_{\Theta}=\left[ 
\begin{array}{cccc}
\theta^{1} & 0 & 0 & 0 \\ 
\theta^{2}+i\theta^{3} & \theta^{8} & 0 & 0 \\ 
\theta^{4}+i\theta^{5} & \theta^{9}+i\theta^{10} & \theta^{13} & 0 \\ 
\theta^{6}+i\theta^{7} & \theta^{11}+i\theta^{12} & \theta^{14}+i\theta^{15}
& \theta^{16}%
\end{array}
\right] .  \label{eq:Cholesky}
\end{equation}
We should keep in mind that while the number of the parameters for the
complex lower triangular matrix (\ref{eq:Cholesky}) is 16, that of the
density matrix (\ref{eq:DM2}) is effectively 15, because of the denominator $%
\mathop{\mathbf{Tr}}\nolimits[T_{\Theta}T_{\Theta}{}^{\dagger}]$.

The coincidence counts $\{N\}=\{n_{\nu}\}_{\nu=1}^{16}$ are assumed to obey
the Poisson distribution with the mean being 
\begin{eqnarray}
M_{\nu}(\Theta) & \equiv & \lambda \mathop{\mathbf{Tr}}\nolimits[%
|m_{\nu}\rangle \langle m_{\nu}|\rho _{\Theta}]  \nonumber \\
&=& \mathop{\mathbf{Tr}}\nolimits[|m_{\nu}\rangle \langle
m_{\nu}|T_{\Theta}T_{\Theta}{}^{\dagger}],  \label{eq:meanPro}
\end{eqnarray}
where we rearrange the parameters in Eq.~(\ref{eq:Cholesky}) so that the
value of $\mathop{\mathbf{Tr}}\nolimits[T_{\Theta}T_{\Theta}{}^{\dagger}]$
coincides with the nuisance parameter $\lambda$ of Eq.~(\ref{eq:nuisance}).
Thus the probability mass function (Poisson density function) of the
measurement results $\{N\}$ for given values of the parameters $\{\Theta \}$
is written as 
\begin{equation}
P(N|\Theta)=\prod_{\nu =1}^{16}e^{-M_{\nu}(\Theta )}\frac{%
M_{\nu}(\Theta)\,^{n_{\nu}}}{n_{\nu}!}\equiv
\prod_{\nu=1}^{16}p(n_{\nu}|\Theta).  \label{eq:pdf}
\end{equation}

Although the parametric model, Eq.~(\ref{eq:DM2}), guarantees the positivity
and trace condition, the simple linear relationship between the results of
measurements $\{N \}$ and the parameters $\{\Theta\}$ like Eq.~(\ref%
{eq:coincidence2}) has disappeared. Nonetheless, the MLE can be applied for
inferring the parameters $\{\Theta \}$ from the observed results $%
\{n_{\nu}\}_{\nu =1}^{16}=\{N\}$ \cite{note_condition}. We can regard Eq.~(%
\ref{eq:pdf}) as a function on the 16-dimensional parameter space where each
point corresponds to a certain quantum state. It is called \textit{%
likelihood function}. Then, it is reasonable to consider that the point
(state) which maximizes the likelihood function (\ref{eq:pdf}) is likely to
be the nearest to the \textit{true} point (state), $\{\theta _{0}^{1},\theta
_{0}^{2},\ldots,\theta_{0}^{16}\}\equiv \{\Theta _{0}\}$. The strategy to
choose the values $\{\hat{\theta ^{1}}(N),\hat{\theta ^{2}}(N),\ldots ,\hat{%
\theta ^{16}}(N)\}$ which maximizes Eq.~(\ref{eq:pdf}) as the estimates is
called maximum likelihood estimation (MLE).

The MLE is elucidated based on the \textit{Kullback-Leibler distance
(relative entropy)} \cite{NCtext2000,CTtext1991,ANtext2000}. It is often
convenient to consider the natural logarithm of the likelihood function $%
P(N|\Theta)$, which is called \textit{log-likelihood function}: 
\begin{equation}
\ln [P(N|\Theta)]=\sum_{\nu =1}^{16}\ln [p(n_{\nu}|\Theta)].
\label{eq:loglikelihood}
\end{equation}
Here it is apparent that this change does not influence the location of the
maximum. As the data acquisition time $t$ is increased infinitely, the
log-likelihood function divided by $t$ tends, with probability 1, to the the
mean log-likelihood function for unit time, $t=1$, i.e., 
\begin{eqnarray}
\frac{1}{t}\ln [P(N|\Theta )] &\approx&
\!\!\sum_{n_{1}=0}^{\infty}\sum_{n_{2}=0}^{\infty}\cdots
\!\!\sum_{n_{16}=0}^{\infty}\!\!P_{0}(N)\ln[P(N|\Theta )]|_{t=1}  \nonumber
\\
&\ & \equiv \mathop{\mathbf{E}}\nolimits_{\Theta_{0}} [\ln[P(N|\Theta )]%
|_{t=1}],  \label{eq:Mloglikelihood}
\end{eqnarray}
where $P_{0}(N)\equiv P(N|\Theta_{0})$ is the \textit{true} probability mass
function of $\{N\}$. The difference between the \textit{true} probability
mass function $P_{0}(N)$ and the parametric model $P(N|\Theta)$ can be
measured by the Kullback-Leibler distance \cite%
{NCtext2000,CTtext1991,ANtext2000}, 
\begin{equation}
D(P_{0}(N)\,\Vert \,P(N|\Theta))=\mathop{\mathbf{E}}\nolimits_{\Theta_{0}}
[\ln [P_{0}(N)]-\ln [P(N|\Theta)]].  \label{eq:relativeE}
\end{equation}
This takes a positive value, unless $P_{0}(N)=P(N|\Theta)$ in all $\{N\}$
(in this case $D(P_{0}(N)\,\Vert \,P(N|\Theta))=0$). Then it becomes clear
that what we try to do by the MLE (i.e., to increase the log-likelihood
function, Eq.~(\ref{eq:Mloglikelihood}), with respect to $\{\Theta\}$) is to
minimize the Kullback-Leibler distance between the \textit{true} probability
mass function $P_{0}(N)$ and its parametric model $P(N|\Theta)$.


\subsection{\label{app:CR bound}Cram\'{e}r-Rao bound and \newline
Fisher information matrix}

The MLE is supposed to be the optimal estimation strategy in the following
sense. The errors of the estimates $\{\hat{\theta ^{1}}(N), \hat{\theta ^{2}}%
(N), \ldots, \hat{\theta ^{16}}(N)\}=\{\hat{\Theta}(N)\}$ can be represented
by the covariance matrix $\mathop{\mathbf{V}}\nolimits(\Theta_{0})=[V^{ij}(%
\Theta_{0})]$ which is given by 
\begin{eqnarray}
&\ & V^{ij}(\Theta_{0}) \\
&=& \mathop{\mathbf{E}}\nolimits_{\Theta_{0}} [(\hat{\theta^{i}}(N) \!-\! %
\mathop{\mathbf{E}}\nolimits_{\Theta_{0}}[\hat{\theta^{i}}(N)])(\hat{%
\theta^{j}}(N) \!-\! \mathop{\mathbf{E}}\nolimits_{\Theta_{0}}[\hat{%
\theta^{j}}(N)])].  \nonumber  \label{eq:covariance}
\end{eqnarray}
The \textit{Cram\'{e}r-Rao inequality} provides an asymptotic lower bound on
the covariance matrix $\mathop{\mathbf{V}}\nolimits(\Theta_{0})$ as follows %
\cite{Helstrom1976,CTtext1991,ANtext2000}. We first assume the \textit{%
unbiasedness} of the estimates $\hat{\theta^{i}}(N)$, i.e., 
\begin{equation}
\mathop{\mathbf{E}}\nolimits_{\Theta_{0}}[(\hat{\theta ^{i}}%
(N)-\theta_{0}^{\,i})]=0.  \label{eq:unbiasedness}
\end{equation}
Then we define the \textit{score} of the probability mass function as 
\begin{equation}
S_{i}(N|\Theta_{0}) \equiv \frac{\partial}{\partial \theta^{i}} \ln[%
P(N|\Theta)]|_{\Theta=\Theta_{0}}=\frac{\frac{\partial}{\partial \theta^{i}}
P(N|\Theta)|_{\Theta=\Theta_{0}}}{P(N|\Theta_{0})}.  \label{eq:score}
\end{equation}
Here, we can readily verify the mean of the score is zero, i.e., 
\begin{equation}
\mathop{\mathbf{E}}\nolimits_{\Theta_{0}} [S_{i}(N|\Theta_{0})]=0.
\label{eq:mean_score}
\end{equation}
Thus, the covariance of the score can be written as 
\begin{equation}
J_{ij}(\Theta_{0})=\mathop{\mathbf{E}}\nolimits_{\Theta_{0}}
[S_{i}(N|\Theta_{0}) S_{j}(N|\Theta_{0})].  \label{eq:Fisher}
\end{equation}
Equivalently, we have 
\begin{equation}
J_{ij}(\Theta_{0})=-\mathop{\mathbf{E}}\nolimits_{\Theta_{0}} [\frac{%
\partial^{2}}{\partial \theta^{i} \partial \theta^{j}} \ln[P(N|\Theta)]%
|_{\Theta=\Theta_{0}}],  \label{eq:Fisher2}
\end{equation}
as can be seen from differentiating Eq.~(\ref{eq:mean_score}) with respect
to $\theta^{j}$. The covariance matrix of the score, $\mathop{\mathbf{J}}%
\nolimits(\Theta_{0}) \equiv [J_{ij}(\Theta_{0})]$, is well known as the 
\textit{Fisher information matrix}. By the \textit{Schwarz inequality} for
expectation, we have 
\begin{eqnarray}
&{\!}& (\mathop{\mathbf{E}}\nolimits_{\Theta_{0}}[%
\sum_{i=1}^{16}z_{i}S_{i}(N|\Theta_{0})\,\sum_{j=1}^{16}y_{j}(\hat{\theta^{j}%
}(N)-\theta _{0}^{\,j})])^{2}  \nonumber \\
&{\le}&
[\sum_{i=1}^{16}\sum_{j=1}^{16}z_{i}z_{j}J_{ij}(\Theta_{0})][\sum_{i=1}^{16}%
\sum_{j=1}^{16}y_{i}y_{j}V^{ij}(\Theta_{0})].  \label{eq:CS}
\end{eqnarray}
where, we introduce two sets of 16 auxiliary real variables $(y_{1}, y_{2},
\ldots, y_{16})\equiv\,^{t}\!\mathop{\mathbf{y}}\nolimits$ and $(z_{1},
z_{2}, \ldots, z_{16})\equiv\,^{t}\!\mathop{\mathbf{z}}\nolimits$. From Eq.~(%
\ref{eq:mean_score}), we have 
\begin{eqnarray}
&{\quad\quad}& \mathop{\mathbf{E}}\nolimits_{\Theta_{0}}[S_{i}(N|\Theta_{0})%
\,(\hat{\theta ^{j}}(N)-\theta _{0}^{\,j})]  \nonumber \\
&=& \mathop{\mathbf{E}}\nolimits_{\Theta_{0}}[S_{i}(N|\Theta_{0})\,\hat{%
\theta ^{j}}(N)]  \nonumber \\
&=&
\!\!\sum_{n_{1}=0}^{\infty}\sum_{n_{2}=0}^{\infty}\cdots\sum_{n_{16}=0}^{%
\infty}\!\! P(N|\Theta_{0})\frac{\frac{\partial}{\partial\theta^{i}}
P(N|\Theta)|_{\Theta=\Theta_{0}}}{P(N|\Theta_{0})} \hat{\theta^{j}}(N) 
\nonumber \\
&=& \frac{\partial}{\partial \theta^{i}} \sum_{n_{1}=0}^{\infty}%
\sum_{n_{2}=0}^{\infty}\cdots \sum_{n_{16}=0}^{\infty}\!\!P(N|\Theta) \hat{%
\theta^{j}}(N)|_{\Theta=\Theta_{0}}  \nonumber \\
&=& \delta^{j}_{i},  \label{eq:CS2}
\end{eqnarray}
where $\delta^{j}_{i}$ is the Kronecker's delta. Consequently, the left-hand
side of the inequality (\ref{eq:CS}) becomes 
\begin{equation}
(\sum_{i=1}^{16}\sum_{i=1}^{16}z_{i}y_{j}\delta^{j}_{i})^{2}=(\,^{t}\!%
\mathop{\mathbf{y}}\nolimits\mathop{\mathbf{z}}\nolimits)^{2}.
\label{eq:lefthand}
\end{equation}
By substituting Eq.~(\ref{eq:lefthand}) and putting $\mathop{\mathbf{z}}%
\nolimits=\mathop{\mathbf{J}}\nolimits(\Theta_{0})^{-1}\mathop{\mathbf{y}}%
\nolimits$ in the Schwarz inequality (\ref{eq:CS}), we obtain 
\begin{equation}
\,^{t}\!\mathop{\mathbf{y}}\nolimits\mathop{\mathbf{J}}\nolimits(%
\Theta_{0})^{-1}\mathop{\mathbf{y}}\nolimits \le \,^{t}\! \mathop{\mathbf{y}}%
\nolimits \mathop{\mathbf{V}}\nolimits(\Theta_{0}) \mathop{\mathbf{y}}
\nolimits ,  \label{eq:CRbound}
\end{equation}
that is, 
\begin{equation}
\mathop{\mathbf{V}}\nolimits(\Theta_{0})\ge \mathop{\mathbf{J}}
\nolimits^{-1}(\Theta_{0}),  \label{eq:CRbound2}
\end{equation}
which is the Cram\'{e}r-Rao inequality for unbiased estimates. Note that
most estimators used in practice are not unbiased. However, the Cram\'{e}%
r-Rao bound on the variance of an unbiased estimator is \textit{%
asymptotically} also a bound on the mean square error, 
\begin{equation}
\mathcal{V}^{ij}(\Theta_{0}) = \mathop{\mathbf{E}}\nolimits_{\Theta_{0}} [(%
\hat{\theta^{i}}(N)-\theta_{0}^{i})(\hat{\theta^{j}}(N)-\theta_{0}^{j})].
\label{eq:MSE}
\end{equation}
of any well-behaved estimator, as shown by Gill and Massar in Ref.~\cite%
{GM2000}. Thus, the Cram\'{e}r-Rao inequality provides us with an \textit{%
asymptotic} lower bound on the covariance matrix $\mathop{\mathbf{V}}%
\nolimits(\Theta_{0})$ for wide variety of estimates in terms of the Fisher
information matrix \cite{Helstrom1976,GM2000,CTtext1991,ANtext2000}. Here,
we mention the significant fact that the maximum likelihood estimates are
asymptotically efficient, in other words, by the MLE, the covariance matrix
asymptotically achieves the Cram\'{e}r-Rao lower bound \cite%
{ANtext2000,Braunstein1992}. In this sense, the MLE is the optimal strategy.




\begin{thebibliography}{99}
\bibitem{note_Bures} The Bures distance $d_{Bures}(\rho_{\Theta_{0}},\rho_{%
\hat{\Theta}})^{\,2}$ does not depend on parametrization. Therefore we can
use any parametrization.
\bibitem{MPinverse} MATHEMATICA 4.0 has a function, \texttt{PseudoInverse},
by which we can calculate the Moore-Penrose generalized inverse of
degenerate matrices.
\bibitem{note_condition} A necessary and the sufficient condition of
tomographic measurements for the linear tomography is not straightforwardly
applied to that for the MLE; nonetheless, we use these condition, because of
its simplicity and the fact that it is at least valid as the sufficient
condition for inferring the parameters by the MLE.
\end{thebibliography}
\end{document}